\documentclass[useAMS]{mn2e}
\usepackage{graphicx,amssymb,amsmath}
\usepackage{hyperref,xcolor}
\usepackage{subfigure}
\usepackage{pdflscape}
\usepackage{mathrsfs}

\title[Unveiling  Vela-SALT ]
  { Unveiling  Vela - Time Variability of Interstellar  Lines in the Direction of the Vela Supernova Remnant II. Na D and Ca II \thanks{Based on
observations obtained with Southern African Large Telescope (SALT) and The Vainu Bappu Telescope (VBT).}}
\author[N. K. Rao et al.  ]
  {N. Kameswara Rao$^1$$^,$$^2$, David L. Lambert$^2$, Arumalla B. S. Reddy$^2$,  Ranjan Gupta$^3$,\\ 
\newauthor
        S. Muneer$^1$ \& Harinder P. Singh$^4$ \\
        \thanks{E-mail: nkrao@iiap.res.in (NKR);
  dll@astro.as.utexas.edu (DLL);muneers@iiap.res.in (SM); rag@iucaa.in (RG); singh@associates.iucaa.in
 (HPS)},\\
  $^1$Indian Institute of Astrophysics, Bangalore 560034, India\\
       $^2$The W.J. McDonald Observatory and Department of Astronomy, The University of Texas, Austin, TX 78712-1083, USA\\
       $^3$Inter University Centre for Astronomy and Astrophysics, IUCAA Pune, 411007 India\\
       $^4$Department of Physics \& Astrophysics, University of Delhi, Delhi 110007 India\\ }
\RequirePackage[normalem]{ulem} 
\RequirePackage{color}\definecolor{RED}{rgb}{1,0,0}\definecolor{BLUE}{rgb}{0,0,1} 
\providecommand{\DIFdelbegin}{} 
\providecommand{\DIFdelend}{} 

\begin{document}

\pagerange{\pageref{firstpage}--\pageref{lastpage}} \pubyear{2014}

\maketitle

\label{firstpage}

\begin{abstract}
    In a survey conducted  between 2011-12
   of  interstellar Na\,{\sc i} D line profiles in the direction of  the Vela supernova remnant,
   a few lines of sight showed
  dramatic changes in  low velocity absorption
 components with respect to profiles from 1993-1994 reported by Cha \& Sembach. 
    Three stars - HD 63578, HD 68217 and HD 76161 showed large decrease in
 strength over the 1993-2012 interval.
   HD 68217 and HD 76161 are associated with the Vela SNR whereas
 HD 63578 is  associated with $\gamma^2$ Velorum
   wind bubble.  Here,
  we present  high spectral resolution observations of Ca\,{\sc ii} K lines obtained
 with the  Southern African Large Telescope (SALT) towards these three stars along with simultaneous observations of Na\,{\sc i} D lines. These
 new spectra confirm  that the Na D interstellar absorption weakened drastically between 1993-1994 and 2011-2012 but show for the first time
 that the Ca II K line is unchanged between 1993-1994 and 2015. This remarkable contrast between the behaviour of Na D and Ca II K line
 absorption lines is a puzzle  concerning gas presumably affected by the outflow from the SNR and the wind from $\gamma^2$ Velorum.

\end{abstract}


\section{Introduction}

                            Vela supernova remnant (SNR) is  the closest relic of
 of a stellar explosion   to earth, located at a distance of 287$\pm$19 pc, as infered from the VLBI parallax
 of its pulsar (Dodson et al 2003) and represents a supernova explosion that
 occurred 11000 years ago (Reichley,  Downs \& Morris1970). Observational studies
 of interstellar lines present in the spectra of stars in the direction of vela
 SNR provide information about the interaction of the remnant with local
 interstellar medium (ISM). Such studies predominantly concentrated on profiles 
  of Ca\,{\sc ii} K  and Na\,{\sc i} D lines superimposed on stellar spectra. 
 
               Recently, we completed a study  of high resolution Na\,{\sc i} D line
  profiles  towards sixty four OB stars in the direction of Vela SNR , 
  mainly carried out during 2011-2012, and compared them with Ca\,{\sc ii} K and Na\,{\sc i} D line profiles of the same stars earlier obtained by Cha \& Sembach (2000) in
  the period 1993-1996 at comparable spectral resolution (Rao et al 2015 -hereafter Paper I).
     Comparison  of the profiles of Na\,{\sc i} D from these two epochs revealed 
  major decreases in low
  velocity absorption components   towards  a few stars. 
  When Paper I was written, we did not have  Ca\,{\sc ii} K profiles   for  comparison
  with  K line observations of Cha \& Sembach (2000) to constrain the cause of this  Na D variability. Subsequently, we
  could obtain  with the high-resolution spectrograph at Southern African Large Telescope (SALT)  Ca\,{\sc ii} K profiles for three of the stars exhibiting dramatic weakening of Na D absorption.
  In the present paper, we report on the Ca\,{\sc ii} K and Na\,{\sc i} D lines towards three stars in which D line
  absorption at low velocities weakened greatly between 1993-1994 and 2011-2012. The surprising result is that  in sharp contrast to the Na D
  lines the Ca\,{\sc ii} K line in 2015 has the same profile in all three cases as it did in 1993-1994.

\DIFdelbegin

\DIFdelend \begin{figure*}
\vspace{0.0cm}
\includegraphics[width=7cm,height=8cm]{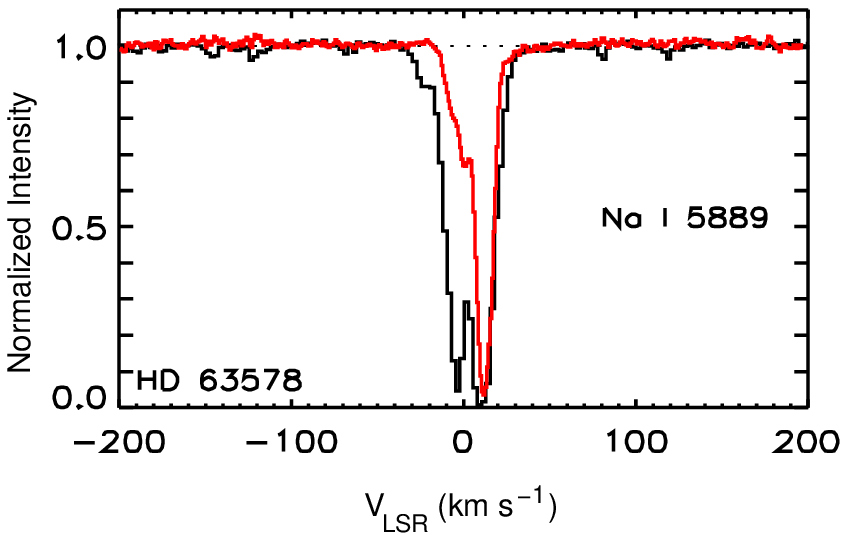}
\vspace{0.0cm}
\includegraphics[width=7cm,height=8cm]{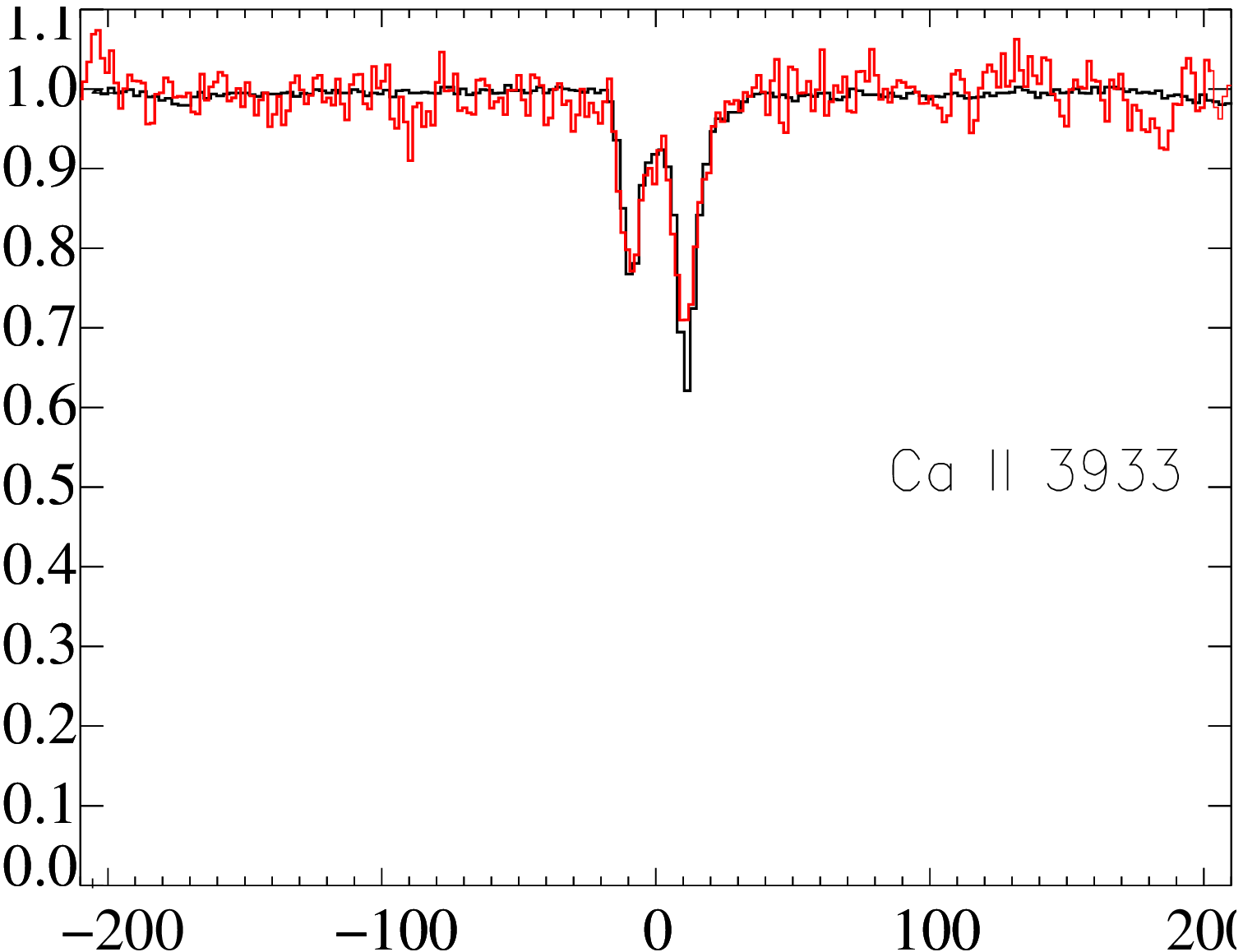}
\caption{(Left):  Profile of  Na\,{\sc i} D$_2$ in HD 63578 obtained in
 2011 (red line) shown superposed on the profile of  Na\,{\sc i} D$_2$ 
 observed in 1993 by Cha \& Sembach (2000)(black line). The blue-shifted
 absorption components in the 1993 spectrum (black line) is absent in the
 2011 spectrum.  (Right): Ca\,{\sc ii} K  profiles of HD 63578 obtained with SALT in
 2015 (red line) overplotted on  profile obtained by Cha \& Sembach (2000) in
1993.}
\end{figure*}


\begin{figure*}
\includegraphics[width=7cm,height=8cm]{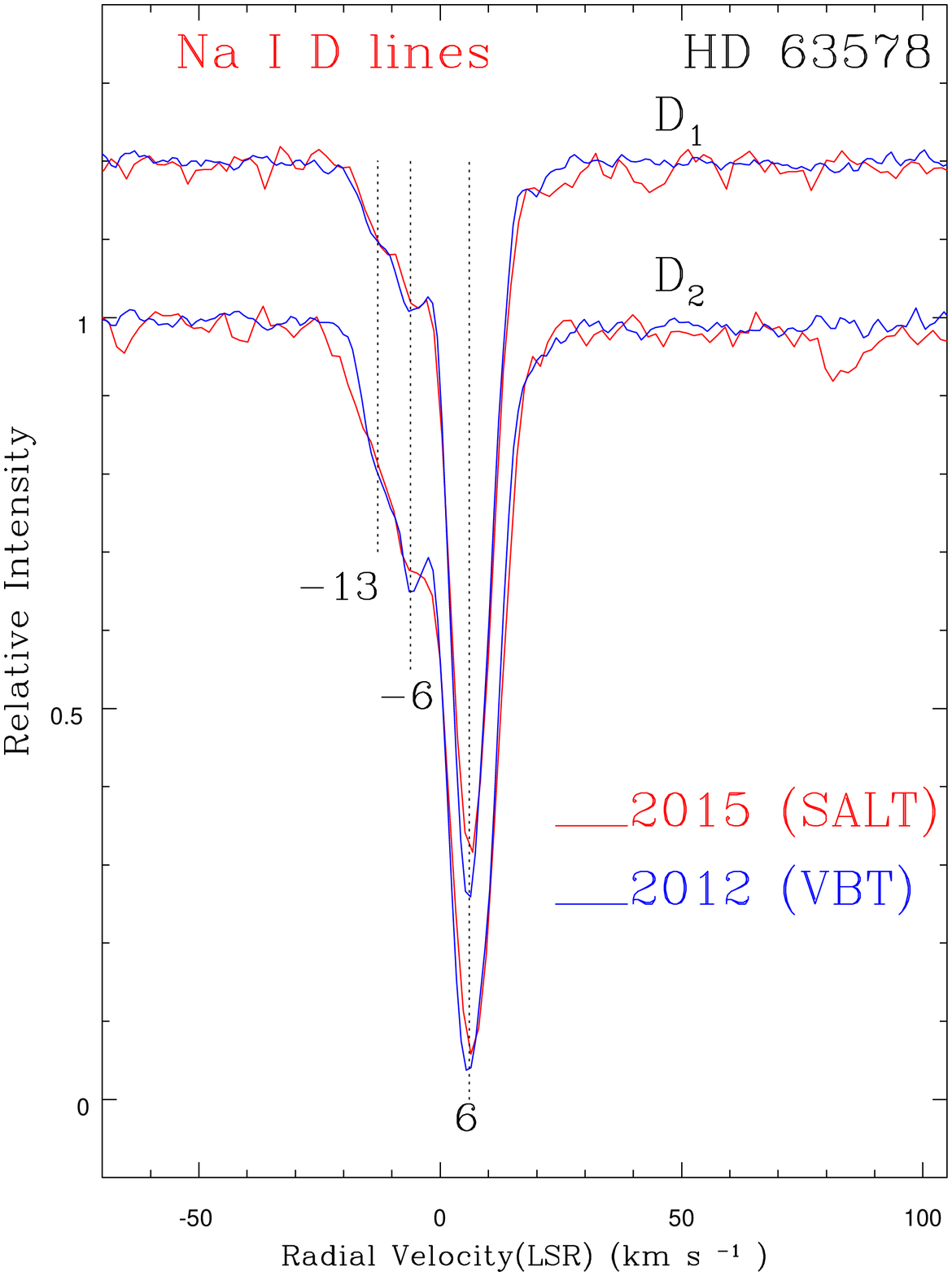}
\caption{ The profiles of Na\,{\sc i} D in  HD 63578 obtained during
 2012 (blue line)   are superposed with  profiles obtained in 2015
 with  the SALT (red line). The SALT spectrum was not corrected
 for   (weak) telluric lines. The dotted lines show the velocities of the four Gaussian components fitted to the
profiles -- see Table 1.}
\end{figure*}

Earlier studies of interstellar absorption lines towards various stars in and around the Vela SNR 
 indicated that high velocity
(i.e $> 100$ km s$^{-1}$) components, most commonly seen  in Ca\,{\sc ii} profiles,
  manifest shocked gas associated with the SNR (Wallerstein,
Silk \& Jenkins 1980; Jenkins, Wallerstein \& Silk 1984).  Sushch, Hnatyk \& Neronov (2011)
 pointed out that only stars with distances greater than 500 pc 
 show  these high-velocity
components in their spectrum (see Paper I figure 18, Cha, Sembach \& Danks 1999).
 For reference, the pulsar is at about 290 pc and the three
stars of interest here are at 300-500 pc and do not show high-velocity interstellar components.  
 The  diameter of Vela supernova remnant is estimated to be about  8.3 degrees (Aschenbach 1993;
Aschenbach, Egger \& Trumper 1995).  
   Two of the three stars discussed here, HD 68217 and HD 76161, are located in the 
outer edge of the ROSAT X-ray image of the SNR. The third one, HD 63758, is located in the $\gamma^{2}$ Vel wind bubble region.

\begin{table*}
\centering
\begin{minipage}{170mm}
\caption{\Large Absorption Lines of Ca\,{\sc ii} K \&  Na\,{\sc i} towards HD 63578 }
\begin{footnotesize}
\begin{tabular}{lcrrrcrrccrrrccr}
\hline
\multicolumn{1}{c}{}&\multicolumn{2}{c}{Ca\,{\sc ii} K }&\multicolumn{1}{c}{}&\multicolumn{1}{c}{}&
\multicolumn{3}{c}{Na\,{\sc i} C \& S}$^a$ &\multicolumn{1}{c}{} & \multicolumn{3}{c}{Na\,{\sc i} (VBT)$^b$}&\multicolumn{1}{c}{}  &\multicolumn{2}{c} {Na\,{\sc i} (SALT)$^c$}  \\
\cline{1-3} \cline{5-8} \cline{10-12} \cline{14-15}  \\
      & C\&S & SALT &  & & &$D_{\rm 2}$&$ D_{\rm 1}$ & &   &$ D_{\rm 2}$& $D_{\rm 1}$&  &$ D_{\rm 2}$& $D_{\rm 1}$ \\
\cline{1-3} \cline{5-9} \cline{10-12} \cline{14-15}  \\
  $V_{\rm LSR}$ & Eq.w &Eq.w&&    & $V_{\rm LSR}$ &Eq.w&Eq.w&  &$V_{\rm LSR}$  &
  Eq.w   & Eq.w &  & Eq.w &Eq.w&  \\
  km s$^{-1}$ &(mA) &(mA) & & &km s$^{-1}$ &(mA)& (mA)& &km s$^{-1}$   &(mA)   &(mA)& & (mA)&(mA)&   \\
\hline
   $-14$   &24  &28   &  &     &   &    &     &   &$-13$   &  31   &15 & &(27)  &14  \\
         &    &     &  &     &$-4$ &223 &178  &   & $-6$   & 39   &24 &  &$^*$  &21   \\
    6    &34  &43   &  &     &   &    &     &   &  6   &232   &191 &  &230 &191  \\
         &    &     &  &     &11 &325 &270  &   &        &      &    & &   &  \\
         &    &     &  &     &   &    &     &   & 22   &   5  &    &  &   &  \\
\hline
\end{tabular}
\\
$^a$ C\&S: Cha \& Sembach (2000) observations from 1993 \\
 $^b$ VBT: Average of four nights from 2011- 2012 (see Paper I)\\
 $^c$ SALT: Observation from 2015 December 15 \\
$^*$ : Affected by telluric line \\

\end{footnotesize}
\label{default}
\end{minipage}
\end{table*}

\section{Observations}

Our high-resolution spectra of the three stars discussed here were obtained at two observatories. The
resolving power $R = \lambda/d\lambda $ of the spectra is comparable to that of the baseline spectra
by Cha \& Sembach (2000) who reported a value of $R = \lambda/d\lambda \simeq 75000 $.

Observations of the Na D lines in Paper I

  have been obtained  with a 45 meter
   fiber-fed
 cross-dispersed echelle spectrometer at the 2.3 meter Vainu Bappu
 Telescope (VBT) at the Vainu Bappu Observatory
(Rao  et al.2005 ).
 $R = \lambda/d\lambda $, the spectral resolving power
 with a 60 $\micron$ m slit, was 72000. The spectrum covers the wavelength range of
4000 to 10000\AA\ with gaps.  Beyond about
5600\AA\  the echelle orders were incompletely recorded on a E2V
2048$\times$4096 CCD chip. Although  Na D lines are captured, the  Ca\,{\sc ii} H \& K lines occur in the insensitive region.
 The wavelength calibration was done using 
 Th-Ar  hollow cathode lamp exposures  that were
  obtained soon after the exposures on the star.

\DIFdelbegin

\DIFdelend \begin{figure*}
\vspace{0.0cm}
\includegraphics[width=7cm,height=8cm]{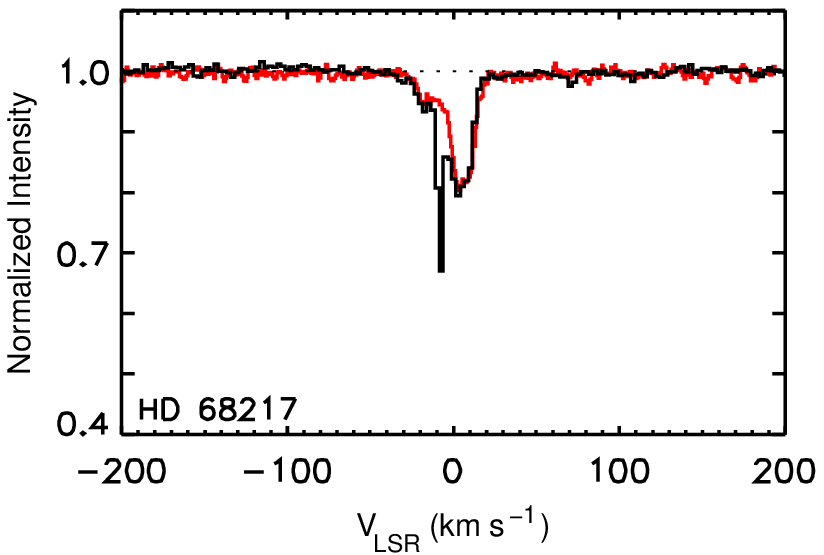}
\vspace{0.0cm}
\includegraphics[width=7cm,height=8cm]{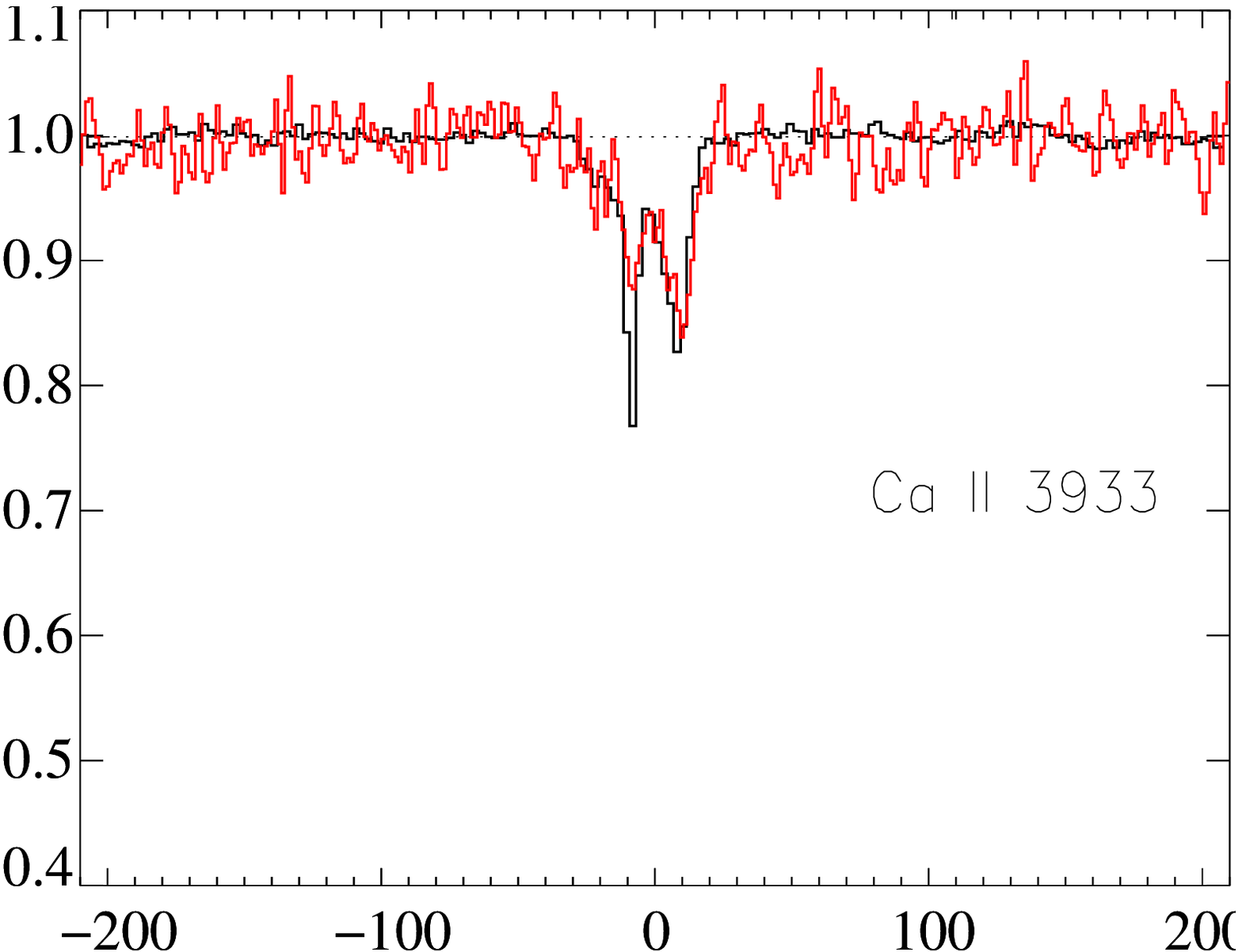}
\caption{(Left panel):Profile of Na\,{\sc i} D$_2$ in the sight line to HD 68217
 as obtained  in 1994 by Cha \& Sembach
 (black line)   superposed on the D$_2$ profile obtained on 2012 January 16
 with the VBT (red line).  The strong absorption
 component present  at $V_{\rm LSR}$ =   $-9$ km s$^{-1}$ in 1994  is conspicuously
   weakened by  2012.
 (Right panel): The Ca\,{\sc ii} K  profile  obtained with SALT in
 2015 of HD 68217(red line) is overplotted on the profile obtained in 1994 by Cha \& Sembach (2000) (black line). There may be a slight weakening of the sharp absorption
 component at $-8$ km s$^{-1}$ by the time the SALT profile was obtained.  }
\end{figure*}

\begin{figure*}
\includegraphics[width=7cm,height=8cm]{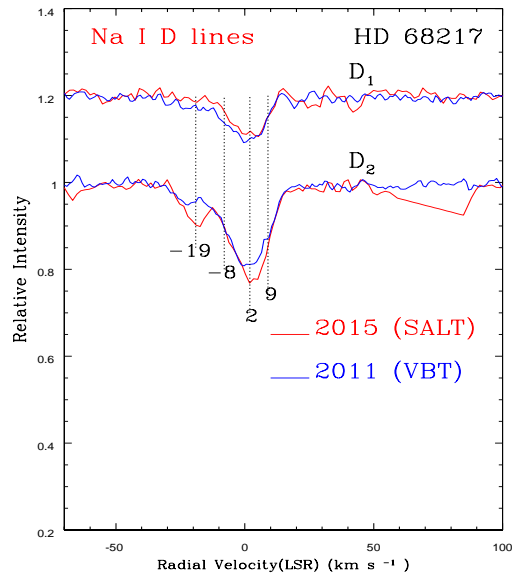}
\caption{  Profiles of  Na\,{\sc i} D$_2$ and D$_1$
 obtained with  VBT  in 2011 (blue lines) are compared with profiles obtained on
 2015 December  25  with SALT (red line). The profiles are very similar indicating no
 significant change occurred between
 2011 and 2015. The SALT spectrum was not corrected
 for (weak) telluric lines.  The dotted lines show the velocities of the Gaussian components fitted to the Na D profiles
-- see Table 2.}
\end{figure*}

\begin{table*}
\centering
\begin{minipage}{170mm}
\caption{\Large Absorption Lines of Ca\,{\sc ii} K \&  Na\,{\sc i} towards HD 68217. }
\begin{footnotesize}
\begin{tabular}{lcrrrcrrccrrrccr}
\hline
\multicolumn{1}{c}{}&\multicolumn{2}{c}{Ca\,{\sc ii} K }&\multicolumn{1}{c}{}&\multicolumn{1}{c}{}&
\multicolumn{3}{c}{Na\,{\sc i} (C \& S)$^a$} &\multicolumn{1}{c}{} & \multicolumn{3}{c}{Na\,{\sc i} (VBT)$^b$}&\multicolumn{1}{c}{}  &\multicolumn{2}{c} {Na\,{\sc i} (SALT)$^c$}  \\
\cline{1-3} \cline{5-8} \cline{10-12} \cline{14-15}  \\
      & C\&S & SALT &  & & &$D_{\rm 2}$&$ D_{\rm 1}$ & &   &$ D_{\rm 2}$& $D_{\rm 1}$&  &$ D_{\rm 2}$& $D_{\rm 1}$ \\
\cline{1-3} \cline{5-9} \cline{10-12} \cline{14-15}  \\
  $V_{\rm LSR}$ & Eq.w &Eq.w&&    & $V_{\rm LSR}$ &Eq.w&Eq.w&  &$V_{\rm LSR}$  &
  Eq.w   & Eq.w &  & Eq.w &Eq.w&  \\
  km s$^{-1}$ &(mA) &(mA) & & &km s$^{-1}$ &(mA)& (mA)& &km s$^{-1}$   &(mA)   &(mA)& & (mA)&(mA)&   \\
\hline
         &    &      &  &     &$-18$ &17 &$\leq$ 6 & &$-19$   & 9  &5 & &$^*$ &3  \\
   $-8.0$ &14  &15    &  &     &$-8$  &30 &20     &   & $-8$   & 10 &5 &  &$^*$ & 3 \\
         &    &      &  &     & 2  &43 &12     &   & 2   & 41   &22 &  &41 &17  \\
     9.0 &20  &25    &  &     & 9  &20 &16     &   & 9   & 21   &10   & & 21 & 9  \\
\hline
\end{tabular}
\\

$^a$ C\&S: Cha \& Sembach (2000) observations from 1994 \\
 $^b$ VBT: Average of four nights from 2011- 2012 (see Paper I)\\
 $^c$ SALT: Observation from 2015 December 15 \\
 $^*$ Affected by telluric line \\

\end{footnotesize}
\label{default}
\end{minipage}
\end{table*}

         New high resolution spectra of Ca\,{\sc ii} H \& K lines along with 
  Na\,{\sc i} D lines have been acquired with the high resolution spectrograph
  (HRS) on the Southern African Large Telescope (SALT) on 2015 December 15 (Bramall
et al. 2010).  The HRS blue spectrum with a 1.6 arc sec fiber provides a resolving power of
   66700. Spectra with Both blue arm (3674-5490 \AA) and red arm (5490-8810 \AA)  
  have been obtained.    The S/N ratio of the red spectrum in the region of the Na D lines is
  comparable to that  of the VBT spectra $\sim 80 - 100)$ and the spectra obtained by Cha \& Sembach (2000).
 However, the SALT spectra around the Ca\,{\sc ii} K line are of a lower S/N ratio ($\sim {35 - 50}$) than the spectra
 provided by Cha \& Sembach (2000).   The resolving power of the SALT spectra is  63000 at
  the D lines as measured from the width of weak telluric lines and comparable at K line.


\DIFdelend \begin{figure*}
\vspace{0.0cm}
\includegraphics[width=7cm,height=8cm]{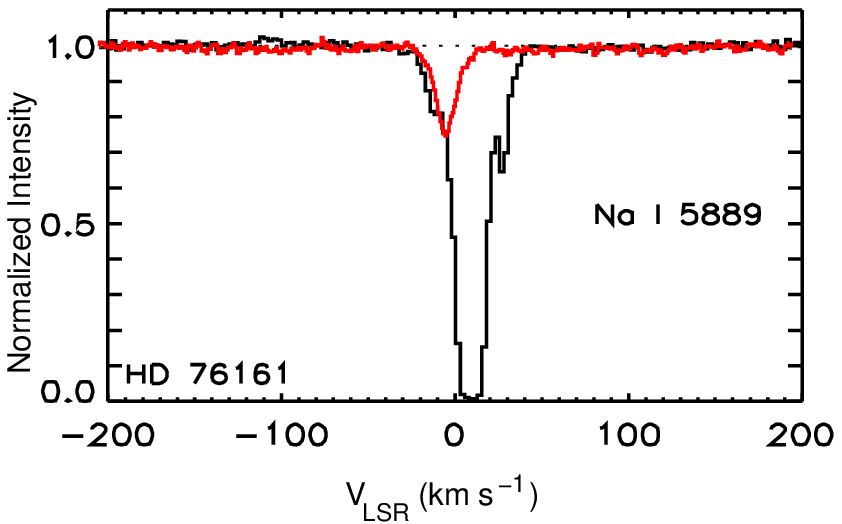}
\vspace{0.0cm}
\includegraphics[width=7cm,height=8cm]{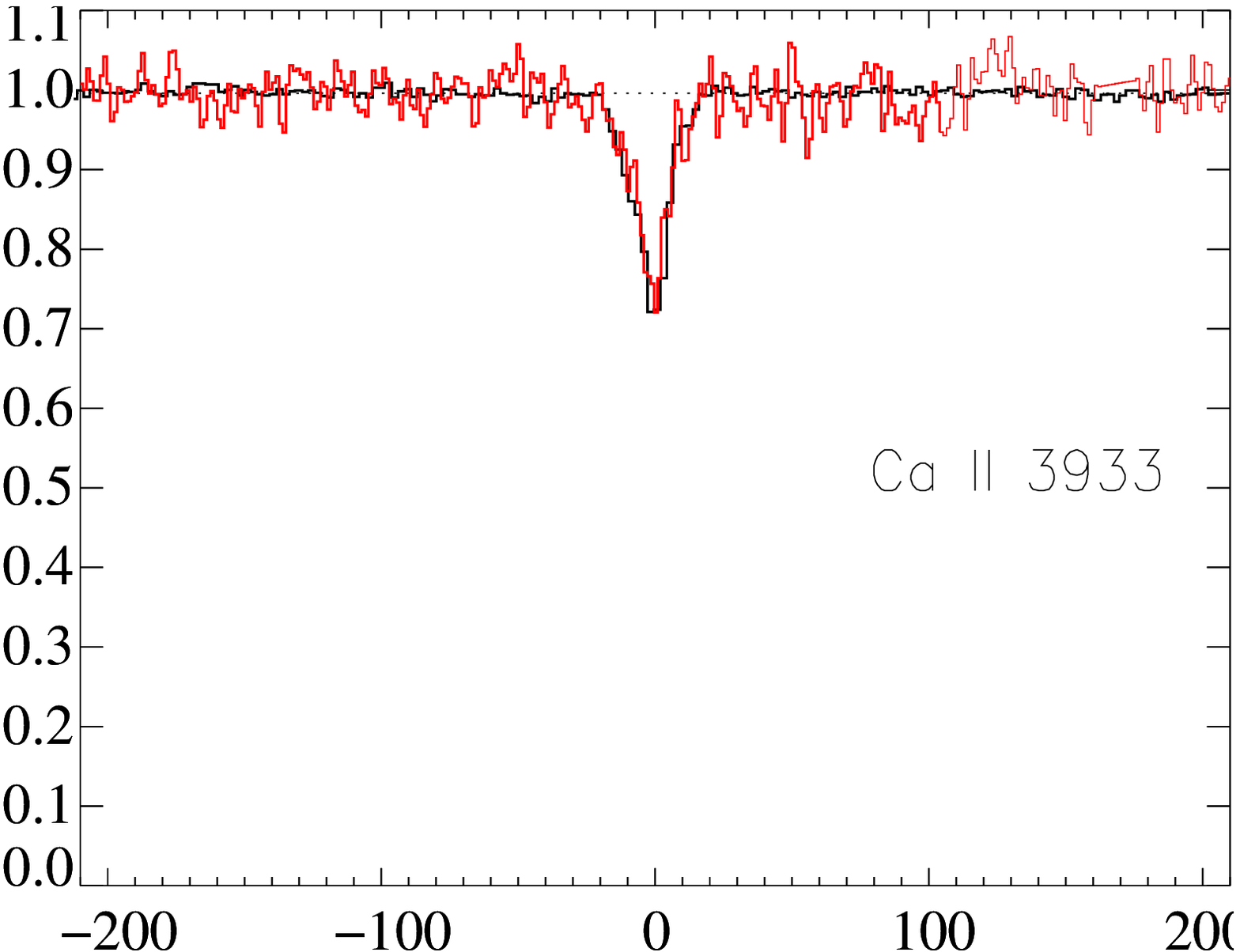}
\caption{(Left panel): Profiles of  Na\,{\sc i} D$_2$ in the sight line towards
  HD 76161 obtained with VBT on 2011 December 25
(red line) is superposed on the 1993 profile obtained by Cha \& Sembach (2000)
 (black line). Strong absorption components at $V_{\rm LSR}$ =   $9$ and $27$
 km s$^{-1}$ have almost disappeared by the time of VBT observations in 2011. 
 (Right panel):   Ca\,{\sc ii} K  profiles of HD 76161 obtained with SALT in
 2015 (red line)  overplotted on 1993 profile obtained earlier by Cha \& Sembach (2000).}
\end{figure*}

\begin{figure*}
\includegraphics[width=7cm,height=8cm]{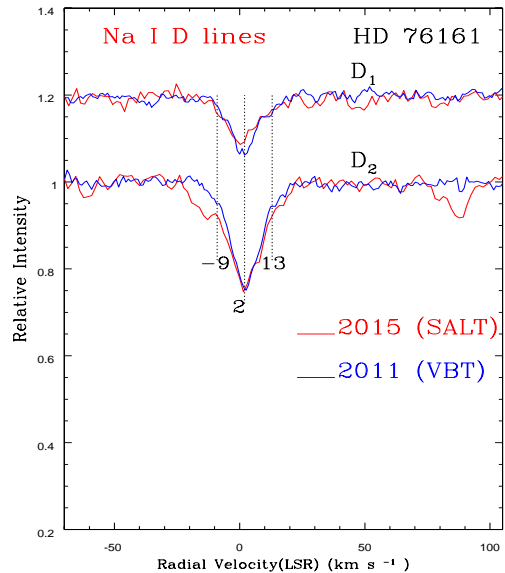}
\caption{  Na\,{\sc i} D$_2$ and D$_1$   profiles
 obtained in 2011 with (blue lines)VBT are compared with profiles obtained on
 2015 Dec 15  with SALT (red line). The profiles are very similar indicating no
 significant change occurred between
 2012 and 2015. The SALT spectrum was not corrected
 for telluric lines. The dotted lines show the velocities of the Gaussian components fitted to the Na D lines -- see Table 3.
}
\end{figure*}

\begin{table*}
\centering
\begin{minipage}{170mm}
\caption{\Large Absorption Lines of Ca\,{\sc ii} K and  Na\,{\sc i} towards HD 76161 }
\begin{footnotesize}
\begin{tabular}{lcrrrcrrccrrrccr}
\hline
\multicolumn{1}{c}{}&\multicolumn{2}{c}{Ca\,{\sc ii} K }&\multicolumn{1}{c}{}&\multicolumn{1}{c}{}&
\multicolumn{3}{c}{Na\,{\sc i} (C \& S)$^a$} &\multicolumn{1}{c}{} & \multicolumn{3}{c}{Na\,{\sc i} (VBT)$^b$}&\multicolumn{1}{c}{}  &\multicolumn{2}{c} {Na\,{\sc i} (SALT)$^c$}  \\
\cline{1-3} \cline{5-8} \cline{10-12} \cline{14-15}  \\
      & C\&S & SALT &  & & &$D_{\rm 2}$&$ D_{\rm 1}$ & &   &$ D_{\rm 2}$& $D_{\rm 1}$&  &$ D_{\rm 2}$& $D_{\rm 1}$ \\
\cline{1-3} \cline{5-9} \cline{10-12} \cline{14-15}  \\
  $V_{\rm LSR}$ & Eq.w &Eq.w&&    & $V_{\rm LSR}$ &Eq.w&Eq.w&  &$V_{\rm LSR}$  &
  Eq.w   & Eq.w &  & Eq.w &Eq.w&  \\
  km s$^{-1}$ &(mA) &(mA) & & &km s$^{-1}$ &(mA)& (mA)& &km s$^{-1}$   &(mA)   &(mA)& & (mA)&(mA)&   \\
\hline
   $-8$    &16  & 6   &  &     &$-6$ & 94 & 32  &   &$-9$    &  6   &3  &  &$^*$ &4  \\
    2    &38  & 43  &  &     &   &    &     &   & 2    & 60   &32 & &63 &28  &   \\
         &    &     &  &     & 9 &395 &365  &   &        &      &   &  & &  \\
         &    &     &  &     &   &    &     &   &13   &  8   & 4  & &10 &6  \\
         &    &     &  &     &27 & 61 & 32  &   &        &      &    &  &   &  \\
\hline
\end{tabular}
\\
$^a$ C\&S: Cha \& Sembach (2000) observations from 1993 \\
 $^b$ VBT: Average of four nights from 2011- 2012 (see Paper I)\\
 $^c$ SALT: Observation from 2015 December 15 \\
$^*$ Affected by telluric line \\

\end{footnotesize}
\label{default}
\end{minipage}
\end{table*}

    IRAF routines have been used for spectral reductions namely flat field 
  corrections, bias subtractions,
  wavelength calibration and corrections for telluric line blending. 
   We adopted the same local standard of rest (LSR) of Cha \& Sembach (2000)
  for converting  heliocentric velocities to LSR velocities.
    Decomposition of various components from the observed line profiles are 
  accomplished by fitting  Gaussian profiles. Central radial velocity
  $V_{\rm LSR} $, full width at half maximum and equivalent width of the 
  components have thus been obtained from these fits. The values of these 
  parameters  listed in Cha \& Sembach (2000) were  used as starting values
  for the gaussian fits and
  further changes were made so as to make  both $D_{2}$ and $D_{1}$ profiles  
  yield  the same number of components, same $V_{\rm LSR} $
  and also similar full width at half maximum. The combined gaussian fits are
  made  to match the observed profiles such that
  no residuals are left over the noise in the surrounding continuum. 
  These fits are of doubtful validity when the lines are saturated, as is the
  case for some Na D profiles.
   When multiple 
 observations are available an average equivalent width of the components is
 presented  in tables. The SALT spectra in the D line region are not corrected
 for telluric line
 contamination, since simultaneous telluric line standard star observations were
 not obtained. As a result radial velocity and equivalent width of few $D_{2}$
 components are affected by telluric lines. However, $D_{1}$ profiles are fine. The LSR radial velocities 
  of the Na\,{\sc i} D line components obtained from SALT profiles agree with in
 $\pm$2 km s$^{-1}$ with the ones measured from VBT profiles. 

\section{Sight lines towards which large variations of Na\,{\sc i} column density
              have been observed }

      Low $V_{\rm LSR} $  velocity  clouds that are recognised through 
  Na D and Ca\,{\sc ii} absorption components  are usually considered to be
  associated with  diffuse interstellar medium in the spiral arms.
  However   interstellar clouds in the proximity of   the Vela supernova
   remnant might have experienced interactions with the remnant through 
   radiation or collisions and thus,
  their physical conditions and kinematics altered.

      Our survey of interstellar  Na D lines (Paper I) showed that the most of the stars
  observed with VBT during 2011-12  had  their Na D profiles  unchanged compared to
 1993-1996 observations of  Cha \& Sembach (2000). 
  Most of the low velocity components, generally in the $V_{\rm LSR}$ range 
  of $\pm25$ to 0 km s$^{-1}$ (Cha, Sembach \& Danks  1999) of  Na D absorption might have their origins in 
 clouds  that occur in  the
 intervening spiral arms.
 A remarkable result of Paper I was a great decrease in strength (almost disappearance)   of low-velocity absorption components in 
 Na D lines in just three stars -- HD 63578, HD 68217 and HD 76161, in the period
 between $1993-1996$
and $2011-2012$.  This trio is discussed here in light of
the SALT spectra of both Na D and the Ca K lines.  Two simple questions may be addressed with the acquisition of
the SALT spectra: (i) Do the 2015 SALT Ca\,{\sc ii} K line profiles differ from
 those that were obtained by  Cha \& Sembach during  1993-1994,  in a similar way to 
the
large changes seen in the Na D profiles between 1993-1994 and 2011-2012? and (ii) Have the Na D profiles evolved further
in the interval three-four year  between the VBT and SALT observations?
(Cha \& Sembach noticed changes in strength and velocity in
 some high velocity absorption components in their observing period of about
 three years, but our VBT
spectra have a limited information on such absorption components. In particular,
 the three stars observed at SALT  have not shown high velocity
components in any available spectra. High-velocity components appeared more frequently in the Ca\,{\sc ii} K line than the Na\,{\sc i}  lines.)

{\bf HD 63578:}  For this star located at a distance of 150 pc behind $\gamma^2$ Velorum and 
seemingly in the bubble generated by
 $\gamma^2$ Velorum's wind ( se Paper I) ,
 Figure 1(left) shows the much weaker low velocity Na D$_2$ line in 2011 relative to that in 1993.
 The figure also shows that the
D$_2$ profile did not change further between 2011 and 2015 (Figure 2) . In striking contrast, the Ca K profiles are unchanged
between 1993 and 2015 (Figure 1-right). 

Gaussian components extracted from the spectra are summarized in Table 1. These show that the Ca K line's two
components are unchanged in velocity and equivalent width between 1993 and 2015 except perhaps for a possible
increase in the redder component in the SALT spectrum. Cha \& Sembach note that the two velocity components
are traceable back to 1971 - 1977. 

At Na D, Cha \& Sembach fitted their 1993 spectrum with  strong components (Table  1) at $-4$ and $+11$ km s$^{-1}$
which differ by several km s$^{-1}$ from their velocities of $-14$ and $+6$ km s$^{-1}$ obtained from the much weaker 
Ca K line.  For the VBT and SALT Na D lines, there are components at $-13$ and $+6$ km s$^{-1}$ apparently coincident
with the Ca K components  plus another weak Na D
component at $-6$ km s$^{-1}$. 

                 Jenkins, Wallerstein \& Silk (1984) obtained an ultraviolet  spectrum of
  the star with the IUE on
  1979 June 15. They measured for both C\,{\sc i},  S\,{\sc ii}, and Si\,{\sc ii}--Fe\,{\sc ii} lines  LSR radial velocities of $+9$, $+7$ and $+2$ km s$^{-1}$, respectively. The $+9$ km s$^{-1}$ velocity is coincident with
   the $+9$ km s$^{-1}$ component of Ca K. The  $-4$ km s$^{-1}$
 component that has been present in 1993 spectrum in Na D lines was not seen in
 either C\,{\sc i} or in singly ionized lines of Ca, S , Si and Fe.
 Thus, this component probably refers to a  cold gas cloud seen  only  in Na D.

{\bf HD 68217:} This star lies near the edge of the ROSAT image of the Vela SNR.  Figure 3(left) shows that a sharp Na D
component at $-8$ km s$^{-1}$ almost vanished by 2011-2012 but remained without additional change in 2015 (Figure 4).  Table 2
shows that the Gaussian components for the Ca K line have close counterparts in the Na D profiles and, in particular,
the sharp feature at $-8$ km s$^{-1}$  which weakened greatly at Na D may also have weakened at Ca K (Figure 3-right).  The four Na D
components identified by Cha \& Sembach are present in the VBT and SALT spectra with the  $-8$ km s$^{-1}$
feature weaker in the latter spectra.  A Na D spectrum obtained in 1989 by Franco (2012) with the same instrument
used by Cha \&  Sembach shows the $-8$ km s$^{-1}$ feature weaker than it was in 1994 but stronger than in VBT and SALT
spectra which suggests that the feature's evolution is not a simple one.

{\bf HD 76161:}   This star which like HD 68217 lies near the edge of the ROSAT image of the Vela SNR (paper I) provides the  third and
the extreme example in which low velocity strong Na D absorptions present in 
  1993 had almost disappeared  by 2011 - see
Figure 5(left) : the saturated absorption component at $V_{\rm LSR} $  $+9 $km s$^{-1}$ has vanished almost completely by 2011.
A redward weaker absorption component at $+27$ km s$^{-1}$ present in 1993 also
 weakened drastically by 2012. 
 However a weak absorption component at about $-9$ km s$^{-1}$ seems
 unchanged over the period of  1993 to 2012 baseline. The Na D profile remained
 unchanged in the period 2011 to 2015 (Figure 6).
 
  The 1993 Ca\,{\sc ii} K profiles  of 1993 and 2015  (Figure 5-right) appear with components at $-8$ and $+2$ km s$^{-1}$ (Table 3).
      The Ca\,{\sc ii} K profile obtained between 1971-77  period by Wallerstein, Silk \& Jenkins (1980) appears to be similar to the profile observed by Cha \& Sembach (2000) in 1993.

                 Jenkins, Wallerstein \& Silk (1984) measured  LSR radial velocity  of S\,{\sc ii} and Si\,{\sc ii}+ Fe\,{\sc ii} as $-3$ km s$^{-1}$ and $-7$ km s$^{-1}$
 respectively from their IUE spectrum obtained in 1979 September 26, very similar 
 values to that of Ca\,{\sc ii} K components measured bu Cha \& Sembach (2000) in 1993.

 Later,  Nichols \& Slavin (2004) detected C\,{\sc i} lines at the LSR velocity of
 8 km s$^{-1}$ in the same IUE spectrum obtained in 1979 . This velocity
 corresponds to the strong Na\,{\sc i} component that disappeared after 1993.
 It appears that a strong neutral cloud  was present at $+9 $km s$^{-1}$ for at least a period
  of 14  years prior to 1993 and  then vanished some time in the next 18 years. 

                  Nichols \& Slavin (2004) investigated C\,{\sc i} lines in 54 stars
  in Vela SNR region including HD 76161, and noted the redshift of C\,{\sc i} lines
  with respect to ionized lines and suggested that C\,{\sc i} lines from the
 ground-state of the atom  are a combination of two absorption components:
  one  near zero velocity which might have its origins in the clouds 
 in the spiral arm  along the sight lines to the star and the other component
 at a higher velocity arising from cloud that might have interacted with
 the Vela SNR. They also suggest that redshift observed in C\,{\sc i} ground-state
 lines might have its origins in a H\,{\sc i} shell on the back side of the 
 remnant.  Dubner et al. (1998) discovered a shell of H\,{\sc i} 
 through 21 cm emission around Vela SNR that follows closely the ROSAT's
  outer X-ray bright shell. This 30 km s$^{-1}$ expanding The H\,{\sc i} shell
  is thought to be a result of recombined postshocked gas behind the advancing
  shock front.
       There is the possibility that the cold neutral cloud 
 that disappeared in the direction of HD 76161 might have been a shocked cloud.

\section{Concluding remarks}

The focus of the paper has been three stars in the direction of the
Vela SNR in which a severe weakening of low-velocity absorption components of
interstellar Na D lines occurred between 1993-1994 and 2011-2012
(Rao et al. 2016). Such low-velocity components are assumed to belong to the
local interstellar medium and not to be an immediate product of the Vela
SNR or other energetic sources in the vicinity of the Vela nebula. 
A few earlier reports for
stars behind the Vela SNR have noted modest variations in the Na D
lines at low velocity - see, for example, observations of the visual binary HD 72127
(Hobbs, Wallerstein \& Hu 1982; Hobbs et al. 1991; Welty. Simon \& Hobbs 2008).
The magnitude of the weakening reported by us earlier is unprecedented in
published studies of interstellar Na D lines (see, for example, McEvoy et al. 2015)
and  is plausibly attributable to
interactions between ambient diffuse interstellar gas and the
high-velocity gas in and around the Vela nebula from the Vela supernova and
the winds from massive stars. It seems highly pertinent to note that out of
 forty one stars with  Na D lines observed previously
 by Cha
\& Sembach  and again at the VBT none showed a comparable strengthening of the Na D lines
between 1993--1994 and 2011--2012.  (As noted for HD 68217, A low-velocity component Na D
did strengthen between 1998 and 1994. Also, HD 73882, a star behind the Vela SNR showed a
strengthening of the Ca\,{\sc i} 4226 \AA\ line between 2006 and 2012 (Galazutdinov et al. 2013)
and presumably the Na D lines also strengthened over this interval.)

The better established signature of these interactions between high-velocity gas and the ambient
interstellar medium
is the presence of high-velocity ($|V_{\rm LSR} \sim 100|$ km s$^{-1}$) absorption
components
seen in stars with distances greater than about 500 pc.
High-velocity
gas was first reported by Wallerstein \& Silk (1971) and Thackeray \&
Warren (1972) from observations of the Ca\,{\sc ii} K line with
extensive follow-up at the Ca\,{\sc ii} K line and Na D lines by Cha \& Sembach (2000).
(High-velocity Na D gas associated with the SNR Monoceros Loop has been shown to be
variable (Dirks \& Meyer 2016).)
High-velocity gas associated with Vela has been studied
in the ultraviolet (Jenkins, Wallerstein \& Silk 1976; Wallerstein,
Silk \& Jenkins 1980; Jenkins et al. 1981; Jenkins, Wallerstein \& Silk
1984). With respect to the Na D and Ca K line, a characteristic of
the high-velocity gas is the several 100-fold increase with respect
to the value for low velocity gas  of the
column density ratio of
Ca\,{\sc ii} to Na\,{\sc i} which is attributed to the destruction of
interstellar grains and release of substantial amounts of Ca (and other
elements) but not Na which is little depleted by grains
(Spitzer 1978; Danks \& Sembach 1995; Sembach \& Danks 1994).

Surely, the outstanding characteristic of the  low-velocity
interstellar lines of the
three sightlines where the Na D lines are strikingly weakened between
1993-1994 and 2012-2015 is that the Ca K line shows no detectable change
between 1993-1994 and 2015 in two cases (HD 63578 and HD 76161) and a possible mild 
weakening in the third case (HD 68217) where the central depth
of the component coincident
in velocity with the varying Na D feature drops from  77\% in
1994 to 88\% in 2015, a drop only slightly larger than the
the amplitude of the noise in the SALT spectrum. In short, the Ca K line is effectively
unaffected by the factors responsible for the large Na D changes and, in
particular, the Ca K's velocity component coincident with the variable
Na D component are unaffected.  This contrast between Na D and Ca K
constrains the factors  driving the weakening of the Na D line.

Estimation of the Na\,{\sc i} column density for two of the three examples is an uncertain procedure
for the 1993-1994 spectra because the dominant components are saturated.
For the VBT and SALT spectra, the variable component of the Na D lines is
little saturated; the equivalent widths of the D$_2$ and D$_1$ lines approach
the 2:1 weak line limit. For HD 68217, the $-8$ km s$^{-1}$ component appears to provide both
the Na D and Ca K line and 
 assuming that the lines are
unsaturated, the column density ratio 
 N(Ca\,{\sc ii})/N(Na\,{\sc i}) was about 3 in 1994 and increased to 17 in 2015. 
For the other two stars, the ratio N(Ca\,{\sc ii})/N(Na\,{\sc i}) for the highly-variable
component was between about two and four in
2015. The 1993-1994 ratios were obviously much smaller because of the larger Na\,{\sc i}
column densities and these ratios appear typical of diffuse clouds (Danks \& Sembach 1995).
 The ratio of
17 for HD 68217 in 2017 is typical of values for high-velocity clouds in the Vela SNR (Sembach \& Danks 1994)
but the change between 1993-1994 and 2015 seems unlikely to be due to release of Ca from grains because the Ca K
line's equivalent width is unchanged over this interval and the increase in the ratio arises from loss of
neutral Na presumably to ionization. High-velocity clouds moving tangentially to the line of sight may appear
as low velocity clouds and this likelihood is increased for stars such as our trio located near the
edge of the SNR.

           Interactions between the SNR and an ambient diffuse diffuse cloud may break the
 cloud into smaller parts. Pakhomov, Chugal \& Iyudon (2012 -- see also Klein, McKee \& Collela 1994)
  discuss how a fast-moving  shock  interacts with a diffuse cloud. Three stages of interaction
  are envisaged. First, the shock    propagates through the cloud. In the second stage, 
 the cloud is accelerated. Finally, the cloud is fragmented. It seems that the interaction can
  destroy a small (diameter of a few au) cloud in a few years. Although detailed
  calculations remain to be done to show that almost complete disappearance of Na D lines may be
   achieved with almost no change of the Ca K line,  it seems, as the referee has pointed out, 
  unlikely that large-scale fragmentation can lead to  a large
  Na D reduction  without a change in Ca K and, moreover, changes in radial velocity are likely to
  accompany column density changes.

             Nonetheless, the fact that these remarkable changes in Na D lines unaccompanied by
   Ca K line changes occur on sight lines  to SNR or through a region crossed by a vigorous 
   stellar wind and have not been seen sight lines through the ambient diffuse stellar medium 
   suggests that shocks may be the key to understanding our remarkable discovery.  Shock fronts
   even at low velocities  can emit Ly-alpha photons (Shull \& McKee 1979).  Lyman $\alpha$ at
   10.2 eV can ionize neutral Na with its ionization potential of 5.14 eV but not Ca$^+$ with
   its ionization potential of 11.9 eV.  As the front approaches the cloud, the  increase in 
  Lyman $\alpha$ flux could be rather abrupt;  the zone containing Lyman $\alpha$
  photons ahead of the front can be narrow because Lyman $\alpha$ photons scatter many times 
  in a neutral   hydrogen medium.   (This appealing argument we owe to the referee.)

            A strictly geometrical explanation but perhaps overlooking the SNR and stellar wind
  association may be possible. The column density ratio of Na D/Ca K for diffuse clouds spans the
  range from 300 to 0.02 (Welty, Morton \& Hobbs 1996). If a thin cloud of high Na D/Ca K moved out
  of the line of sight to our three stars, the Na D line would weaken sharply but, perhaps, the
  Ca K line would be largely unchanged.  Observations of sight lines through the general
  interstellar medium do not support this idea as a common an event as we find towards the Vela SNR.

The principal next challenge suggested by our VBT and SALT spectroscopy of 
Na D and Ca K lines is to catch and follow
large scale changes occurring along a line of sight.
 Panoramic optical spectroscopy will be able to observe the behaviour of
many interstellar lines (see, for example, Pakhomov, Chugai \& Iyudin 2012).   
Since there is as yet
 no predictor of which star is about to undergo a change in its
interstellar lines, detection of an onset will require routine high-resolution
monitoring of
a set of stars in and around the Vela SNR but this presents a severe scheduling
challenge and involves a considerable consumption of observing time.
Although observation of transient phenomena is a `hot'
contemporary topic, its remit does not yet to extend to exploration of
the questions raised by our paper.

\section{Acknowledgments}

We thank SALT astronomer, Brent Miszalski, for his considerable help in conducting the HRS 
 observations for us. We also thank Dr. Ed Jenkins, the referee, for his insightful comments 
 about how shocks may account for our results and other corrections.
 We also would like to thank Baba Varghese for 
providing the figures of Ca\,{\sc ii} K superpositions. 
 This research has made use of the SIMBAD database, operated
at CDS, Strasbourg, France.  We also would like to thank 
 staff of  the VBO at  Kavalur for their help with observations.

\end{document}